\documentstyle[12pt,psfig]{article}

\begin{document}
\begin{flushleft}
{J\"ulich Preprint IKP-Th 25/1995 \hfill October 11, 1995 } \\
{Cavendish Preprint HEP 95/3 (revised) \hfill  } \\
\end{flushleft}
\vspace{1.0cm}
\begin{center}
{\huge \bf The final state hadrons in polarised deep inelastic scattering
\\ }
\vspace{3ex}
\vspace{3ex}
{\large \bf S. D. Bass \\}
\vspace{3ex}
{\it Cavendish Laboratory,
University of Cambridge, Madingley Road, Cambridge CB3 0HE, U.K. \\}
\vspace{3ex}
{\it and \\}
\vspace{3ex}
{\it Institut f\"ur Kernphysik, KFA-J\"ulich, D-52425 J\"ulich, Germany
\footnote{Present address} \\}
\vspace{3ex}
{\large \bf Abstract \\}
\end{center}

General arguments suggest that the non-perturbative background field in QCD
may have a non-trivial
spin structure.
We discuss how this effect may be manifest in semi-inclusive measurements of
fast pions
in polarised deep inelastic scattering.

\vspace{2.0cm}

\section {Introduction }

The discovery of the EMC spin effect [1] has inspired a vigorous programme
to understand
the internal spin structure of the nucleon -- for reviews see [2,3].
In the absence of a solution to non-perturbative QCD we construct QCD inspired
models of the nucleon
where the valence quarks move
in some background field that is defined self consistently as the deformation
of
the non-perturbative vacuum that is induced by the valence quarks.
The background field includes the effects of confinement and the dynamical
breaking of
chiral symmetry.
It is modelled by the bag in relativistic quark models of nucleon structure.
One of the key questions in QCD spin physics is {\it does this background field
possess a non-trivial
spin structure ? }
This problem has been addressed by a number of authors [4-9].
(It is important to distinguish polarised gluons inside the bag and a possible
spin
structure of the long-range gluon induced fields which are described by the bag
itself.)

Nachtmann and collaborators [4] have proposed that the quark sea may develop a
nett
transverse polarisation
as the
quarks move through the non-perturbative
colour-magnetic fields in the QCD vacuum similar to the build
up of electron polarisation in storage rings.
This effect could lead to a breakdown of factorisation in polarised Drell-Yan
experiments
due to initial state correlations
where the
anti-quark of one hadron moves through the non-perturbative vacuum of the
second.
This effect is consistent with the NA-10 data [10]
and offers a possible explanation of the K-factor in Drell-Yan experiments [4].

The question of background field effects in polarised deep inelastic scattering
has
been addressed
by Jaffe and Manohar [5,9], Fritzsch [6] and the present author [7,8].
Jaffe and Manohar [9]
showed that it is necessary to consider non-perturbative gluon fields in order
to
fully  understand
the role of the anomaly in the first moment of $g_1$.
This approach was also taken by Fritzsch [6] who proposed a model where the
anomaly
induces two gluonic contributions
to $g_1$:
the partonic contribution from the polarised gluon distribution that was
initially found
by Efremov
and Teryaev and by Altarelli and Ross [11], and also a non-perturbative
contribution
which is associated with the internal structure of the constituent quark.
Bass [7,8] generalised the arguments of Jaffe and Manohar to the higher moments
of
$g_1$
and suggested that the anomaly has the potential to contribute an OZI
violation to $g_1$ at large $x$ ($x$ greater than 0.2)
in the form of a local interaction between the hard photon and the background
field.

\section {Three reasons to look for a large $x$ anomaly contribution to $g_1$ }

Before discussing the phenomenology of a possible large $x$ anomaly in
semi-inclusive
reactions
it is helpful
to keep in mind the reasons why one might
expect to see such an effect
which does not appear in perturbation theory.
We briefly discuss three reasons why one might look for a large $x$ anomaly in
$g_1$.
In this section
we collect together ideas that are scattered through-out the literature and
discuss
the relationship between a possible large $x$ anomaly and quark model
calculations of
$g_1$.
We start by defining the spin dependent quark distribution $\Delta q (x,Q^2)$
via the light-cone
expansion so that:
\begin{equation}
2M s_+ (p_+)^{2n} \int^1_0 dx \ x^{2n} \Delta q_{k} (x, Q^2) =
<p,s | [ {\overline q}(0) \gamma_+ \gamma_5 (i D_+)^{2n} {\lambda^k \over 2}
q(0) ]_{Q^2}^{GI} |p,s >_c,
\end{equation}
where
$D_{\mu} = \partial_{\mu} + ig A_{\mu} $ is the QCD-gauge covariant derivative
and
the superscript $GI$
emphasises that we are using gauge invariant operators.

\begin{enumerate}
\item
Perhaps the most illuminating derivation of the anomaly is due to Schwinger
[12]
and uses
point splitting regularisation (see also [13]).
One finds that the chirality of a quark propagating in some background gauge
field is
not conserved
with the result that
the background field screens the spin of the quark.

We can write $\Delta q(x, Q^2)$ in terms of a light-cone correlation function
in
$A_+=0$
gauge
as the Fourier transform of the axial vector current point-split along the
light-cone:
\begin{equation}
\Delta q (x,Q^2) =
{1 \over 2 \pi} \int dz_- \cos (x z_- p_+)
<p,s| [{\overline q}(z_-) \gamma_+ \gamma_5 q(0)]_{Q^2} | p, s>_c
\end{equation}
The Fourier transform in equ.(2)
means that
small $z_-$ effects have the potential to contribute over a complete range of
$x$ and that large $z_-$
effects
contribute only at small $x$.

Light-cone correlation functions need to be treated with care in an interacting
theory
like QCD
because of ultra-violet divergences [14]. (Each moment of equ.(2) has to be
renormalised
separately.)
Quark and lattice models are endowed with an implicit or explicit ultra-violet
cut-off
so that the correlation functions are formally well defined within such models.
This result has led to suggestions
that
one can use the correlation functions to calculate the
leading
twist part
of the structure function within one's favourite model at some low scale
[15,16].
(The model distributions are then evolved to higher $Q^2$, convoluted with the
QCD radiative
coefficients
and compared with data.)
There is one important practical difference between the correlation function
for
$\Delta q(x,Q^2)$
in interacting QCD and in these models:
the treatment of the $z_- \rightarrow 0$ limit of the point-split matrix
element
in equ.(2).
This limit is where one finds the axial anomaly in Schwinger's derivation of
the
anomaly in the axial-vector current [12] -- see also [8].
The way that we treat
the
$z_- \rightarrow 0$ limit
determines the symmetry which is inherent in the polarised quark
distribution.
At this point one needs to consider QCD directly.
The ultra-violet cut-off in the models means that the model correlation
functions do
not include this zero
light-cone correlation length effect
which, via the Fourier transform,
has the potential to be important at all $x$.

\item

In classical field theory the axial vector current is conserved and defines a
good
spin operator.
In renormalised QCD one finds that the conserved axial vector current differs
from
the physical
gauge-invariant operator by a gauge dependent counterterm, which is commonly
denoted
$k_{\mu}$.
In general gauges, eg. the covariant Feynman gauge, one finds that the forward
matrix
element
of this $k_{\mu}$ are invariant under ``small" gauge transformations of
perturbation
theory
but are not invariant under ``large" gauge transformations with non-zero
winding number
[9].
The parton model is nearly always formulated
in the light-cone gauge $A_+=0$.
In this gauge one finds that
the forward matrix elements of the $k_{\mu}$ are invariant under both small and
large
gauge
transformations
and also that $k_{\mu}$ corresponds with the gluon spin operator (evaluated in
this
gauge).
For this reason, $k_{\mu}$ is commonly associated with the polarised gluon
distribution $\Delta g(x, Q^2)$
in the parton model. ($\Delta g(x, Q^2)$ is a small $x$ effect.)
When one considers the effect of the anomaly in the whole distribution rather
than just
the first moment,
one finds a gauge dependent gluonic
counterterm for each of the $C=+1$ axial tensor operators in equ.(1).
These higher moment anomalous contributions are similarly associated
with $\Delta g (x,Q^2)$
in the parton model.
The interesting question is then:
{\it what is the physics content of the remaining quark distribution ?}
Naively,
it is tempting to say that since this distribution corresponds to a conserved
axial
vector current
it should
therefore
correspond to the same canonical physics that we expect from semi-classical
quark
models.
However, this is not clear in view of the general problem of invariance under
large
gauge transformations.
In a covariant gauge one can continue to make large gauge transformations,
going around
the circle enough times,
so that the anomalous ``$k_{\mu}$ contribution" to equ.(1) becomes arbitrarily
big at
any given $x$.
Given our present understanding of non-perturbative QCD, there is no reason to
assume
that the nett anomalous contribution to the physical
$\Delta q (x, Q^2)$ is zero at large $x$ (say at $x$ greater than 0.2).
If there is a nett anomalous contribution to $g_1$ at large $x$
it would be manifest as an OZI violation
and would be
wrapped up in what we call a constituent quark
in
the light-cone gauge.
In this parton model gauge we write for the first moment of $\Delta q (x,Q^2)$
\begin{equation}
\Delta q^{GI}(Q^2)
= (\Delta q_{S} + \Gamma) - \Biggl( {\alpha_s \over 2\pi} \Delta g \Biggr)
(Q^2)
\end{equation}
Here
$- {\alpha_s \over 2 \pi} \Delta g$
is the partonic polarised gluon contribution [11]
and
$\Gamma$ is a background field contribution which includes the remaining OZI
violation.
The
$\Delta q_{S}$ are the quark ``spin fractions" with good OZI and correspond to
the
quantities
that one would calculate in a semi-classical quark model with no anomaly.
The background field contribution is not present in perturbation theory.
It is present in the spin dependent quark distribution, which is a
non-perturbative
quantity,
so that the sum $(\Delta q_S + \Gamma)$
is the polarised ``quark contribution" to the first moment of $g_1$ in the
parton model.
Indeed, the background
field contribution
has the same $Q^2$ evolution equation and the same local coupling to the hard
photon as we expect of a ``quark" [7].
Equ.(3) is implicit in the model of Fritzsch [6].

\item

It is well known that the current $k_{\mu}$ couples to a massless
(Kogut-Susskind)
ghost pole [17],
which is exactly cancelled by the massless pole that couples to the conserved
(gauge dependent) axial-vector current.
Let us temporarily consider the effect of $k_{\mu}$ in isolation to the physics
of
the gauge invariant current
-- that is,
in the spirit of the parton model neglecting a possible finite non-perturbative
background field
contribution.
In this case, it is natural to ask what is the contribution of the (unphysical)
ghost
to the shape of $g_1$.
Here one employs the Sullivan mechanism [18]
to calculate mesonic contributions to structure functions.
The zero mass ghost would give a contribution to $g_1$ at much larger $x$ than
the light
mass pion.
Of course, this contribution is unphysical and is cancelled by the ghost
coupling to the
conserved current.
However, it does illustrate the point that one cannot self-consistently treat
the
physics
of ``the anomaly" (as described by $k_{\mu}$) in isolation to the conserved
current as a
small $x$ effect.

\end{enumerate}

The correct way to describe the $x$ dependence of the anomaly in $g_1$ would be
an all
moment
generalisation of the analysis of Shore and Veneziano [19], where the OZI
violations
are isolated explicitly.
However, it is not clear
(at
least to the author)
how one would estimate the size of the effect.
At the present time one can only say that non-perturbative QCD appears to
permit an
anomalous
contribution
to $g_1$ at large $x$.
One should therefore try to observe this effect in the laboratory.
This may be possible in semi-inclusive measurements of fast pions in polarised
deep
inelastic scattering --
experiments which are being carried out as part of the HERMES [20]
and SMC [21] programme.
Before we discuss how the anomaly may show up
in these experiments,
it is worth spending time to discuss whether this effect consistent with -- or
indeed is ruled out by --
existing data.
We consider only the inclusive spin asymmetries.
Measurements of semi-inclusive asymmetries are just beginning [21] and
the errors are presently
much too large to make definite conclusions.

It is sometimes emphasised that valence quark models
predicted the inclusive longitudinal
spin asymmetry (and hence $g_1$)
at
$x > 0.2$ within the present errors
[22-24] -- see also [25].
Is a large $x$ anomaly consistent with these calculations ?
The answer to this question is yes.
In the picture we are presenting, the anomaly emerges as a non-trivial spin
structure
in the transition
from the constituent to current quark.
The valence quark model calculations of $g_1$ employ either a Melosh
transformation or
phenomenological
``spin dilution factors" to transform from constituent quark to current quark
degrees of
freedom.
Suppose that we compare $\Delta q (x, Q^2)$ and the C-odd, anomaly free,
polarised
valence distribution
\begin{equation}
\Delta q_V (x,Q^2) =
(q - {\overline q})^{\uparrow}_{GI} - (q - {\overline q})^{\downarrow}_{GI}
=
(q - {\overline q})^{\uparrow}_S    - (q - {\overline q})^{\downarrow}_S.
\end{equation}
A large $x$ anomaly in $\Delta q (x, Q^2)$ would be manifest within these
models as
a different choice of the
parameters in the spin dilution factors
for each of $\Delta q_V (x, Q^2)$ and $\Delta q (x, Q^2)$.
Given the present experimental error on the large $x$ data points, there is
already
plenty of room to vary these parameters and still provide a good fit to the
measured
asymmetries [25].
In our picture the ``spin" carried by the $q_S$ is conserved. If one applies
Schwinger's
derivation
of the anomaly to each of the valence quarks in the model wavefunction, it
follows that
the
screening effect is
proportional to the valence quarks' spin.
It follows that the leading large $x$ behaviour of $g_1 \sim (1-x)^3$ that
follows from
the counting
rules [26]
should not be affected by the presence (or otherwise) of $\Gamma$.
(It is interesting to note that the phenomenological parametrisation of the
anomaly
obtained in [27]
has the correct leading $(1-x)^3$ behaviour.)

\section { How to look for a large $x$ anomaly in semi-inclusive measurements
of
polarised deep
inelastic scattering }

In an ideal world one would like to measure the C-odd spin structure function
directly
and compare
the C-even and C-odd polarised quark distributions at large $x$.
This experiment would involve a neutrino beam and a polarised target and is
clearly
impracticable
at the present time.
The best available tool is to reconstruct the valence distributions from
semi-inclusive
measurements of fast pions in polarised deep
inelastic scattering.
In the rest of this paper we discuss how a large $x$ anomaly should show up in
these
experiments.
We shall concentrate only on the region $x > 0.2$ where polarised gluons, the
Dirac sea
and mesonic effects are not important.

If we assume a large $x$ anomaly then
\begin{equation}
g_1 |_{x > 0.2}
= {2 \over 9} (u^{\uparrow} - u^{\downarrow})_S + {1 \over 18} (d^{\uparrow} -
d^{\downarrow})_S + {1 \over 3} \Gamma
\end{equation}
where we work within flavour SU(3).
Given that the $q_S$ are sufficient to describe the physics of the C-odd
structure
function
and also the local coupling of $\Gamma$ to the hard photon [7], we shall treat
$\Gamma$
as a new ``parton", unique to $g_1$,
where ``parton" is defined according to Feynman [28].

Given our ``new parton", we can modify the naive parton model analysis of
semi-inclusive
deep inelastic scattering [29,30]
to include the background field contribution.
We shall follow the notation of Frankfurt et al. [29].
The $q_S$ fragment
to a fast pion in the same way as the naive parton model quark -- independent
of its
helicity.
We let $z$ denote the fraction of the hard photon's energy which is
taken by
the fast pion.
Following [29],
we use $D_1(z)$
to denote the favoured fragmentation
\begin{equation}
D_1(z) \equiv D_{u_{S}}^{\pi^+}(z) = D_{d_{S}}^{\pi^-}(z)
= D_{\overline{d}_{S}}^{\pi^+}(z) = D_{\overline{u}_{S}}^{\pi^-}(z)
\end{equation}
and $D_2(z)$ to denote the unfavoured fragmentation function
\begin{equation}
D_2(z) \equiv D_{d_{S}}^{\pi^+}(z) = D_{u_{S}}^{\pi^-}(z)
= D_{\overline{u}_{S}}^{\pi^+}(z) = D_{\overline{d}_{S}}^{\pi^-}(z).
\end{equation}
The strange quark fragmentation function
\begin{equation}
D_3(z) \equiv D_{s_{S}}^{\pi^+}(z) = D_{s_{S}}^{\pi^-}(z)
= D_{\overline{s}_{S}}^{\pi^+}(z) = D_{\overline{s}_{S}}^{\pi^-}(z)
\end{equation}
is not relevant to our analysis since we are considering  only the large $x$
region $x > 0.2$
which is not sensitive to the strange quark components in the Fock expansion
of the
nucleon wavefunction.

In order to understand how an explicit background field contribution should
fragment
into fast pions in the final state it is important to note that $\Gamma$ does
not
itself
have any Fock components so that its fragmentation is constrained by the
fragmentation
of the valence quarks.
(For the same reason, since the background field is a property of the vacuum
-- or bag itself --
its contribution to $g_1$ is independent of the number of accessible flavours.)
We give $\Gamma$
its own fragmentation function $D_4 (z)$.
(We make
the assumption here
that $q_S$ and $\Gamma$ can separately be treated in impulse approximation and
that
the fragmentation of our $\Gamma$ parton into fast pions factorises.)
Since $\Gamma$ is a flavour singlet
effect
it will fragment equally into fast $\pi^+$ and $\pi^-$ in the final state,
whence
$D_4$
must be proportional to the sum of
the favoured and unfavoured light-quark fragmentation functions
$D_1$ and $D_2$.
(The fragmentation properties of the anomaly were also discussed in
ref. [31],
where it was assumed that the anomaly can be associated entirely with the
polarised gluon
distribution $\Delta g(x,Q^2)$ -- that is, a purely small $x$ effect.)

We let $N_{\uparrow \Downarrow}^{\pi^+}(x,z)$ denote the number of $\pi^+$
produced in a bin characterised by Bjorken x ($>0.2$) and z,
where the virtual photon helicity is $\uparrow$ and the target proton helicity
is $\Downarrow$.
The spin averaged pion production rates are independent of the anomaly.
It follows that:
$$
N_{\uparrow \Downarrow}^{\pi^+} \sim
{4 \over 9} u_S^{\uparrow}(x) D_1(z) + {1 \over 9} d_S^{\uparrow}(x)
D_2(z)
+ {1 \over 12} \Gamma (x) D_4(z)
$$
$$
N_{\uparrow \Uparrow}^{\pi^+} \sim
{4 \over 9} u_S^{\downarrow}(x) D_1(z)
+ {1 \over 9} d_S^{\downarrow}(x) D_2(z)
- {1 \over 12} \Gamma (x) D_4(z)
$$
$$
N_{\uparrow \Downarrow}^{\pi^-} \sim
{4 \over 9} u_S^{\uparrow}(x) D_2(z)
+ {1 \over 9} d_S^{\uparrow}(x) D_1(z)
+ {1 \over 12} \Gamma (x) D_4(z)
$$
$$
N_{\uparrow \Uparrow}^{\pi^-} \sim
{4 \over 9} u_S^{\downarrow}(x) D_2(z)
+ {1 \over 9} d_S^{\downarrow}(x) D_1(z)
- {1 \over 12} \Gamma (x) D_4(z)
\eqno(9)
$$
whence
$$
N_{\uparrow \Downarrow -\uparrow \Uparrow}^{\pi^+ - \pi^-} \sim
\biggl[{4 \over 9}(u_S^{\uparrow}-u_S^{\downarrow})
     - {1 \over 9}(d_S^{\uparrow}-d_S^{\downarrow})\biggr] (D_1 - D_2)(z)
$$
$$
N_{\uparrow \Downarrow + \uparrow \Uparrow}^{\pi^+ - \pi^-} \sim
\biggl[{4 \over 9}(u_S^{\uparrow}+u_S^{\downarrow})(x)
     - {1 \over 9}(d_S^{\uparrow}+d_S^{\downarrow})(x)\biggr] (D_1 - D_2)(z)
$$
$$
N_{\uparrow \Downarrow -\uparrow \Uparrow}^{\pi^+ + \pi^-} \sim
\biggl[{4 \over 9}(u_S^{\uparrow}-u_S^{\downarrow})(x)
     + {1 \over 9}(d_S^{\uparrow}-d_S^{\downarrow})(x)\biggr] (D_1 + D_2)(z)
     + {1 \over 3} \Gamma (x) D_4(z)
$$
$$
N_{\uparrow \Downarrow + \uparrow \Uparrow}^{\pi^+ + \pi^-} \sim
\biggl[{4 \over 9}(u_S^{\uparrow}+u_S^{\downarrow})(x)
     + {1 \over 9}(d_S^{\uparrow}+d_S^{\downarrow})(x)\biggr]
(D_1 + D_2)(z)
\eqno(10)
$$
It follows from equ.(10) that the spin asymmetry $A^{\pi^+ - \pi^-}$
measures the C-odd valence part of $g_1^N$ in QCD; viz.
$$
A^{\pi^+ - \pi^-}_p = {4\Delta u_v - \Delta d_v \over 4u_v - d_v}
\eqno(11)
$$
for a proton target
and (modulo nuclear effects [32])
$$
A^{\pi^+ - \pi^-}_d = {\Delta u_v + \Delta d_v \over u_v + d_v}
\eqno(12)
$$
for a deuteron target.
A measurement of these asymmetries will allow us to extract the valence
spin distributions $\Delta u_v(x)$ and
$\Delta d_v(x)$.
Note that equs.(11,12) are the same expressions that one also obtains in the
naive parton model
with no background field contribution [29,30].
However,
these C-odd distributions are not in general equal to the C-even distributions
which describe the
inclusive structure function $g_1$.
It is a challenge for future experiments to check whether the parton
distributions
that one uses to describe $g_1$ [33]
can also be used to describe these semi-inclusive C-odd asymmetries.
This will require high quality data with much reduced errors in the
large $x$ region.
If different parton distributions are required to describe both $g_1$ and
also the C-odd semi-inclusive asymmetries
then we would evidence for a large $x$ anomaly.
(Note that if one of our impulse or factorisation hypotheses were to fail, then
it
would be most unlikely that
our parton distributions would describe both $g_1$ and equs.(11,12).
In this respect, these assumptions are not necessary. They are necessary if one
finds
a large $x$ anomaly and
then wishes to make predictions for other processes.)

In this analysis, the spin asymmetry which is obtained by summing over fast
$\pi^+$ and
$\pi^-$ in the final state
can be used to deduce the fragmentation function for the large $x$ part
of the anomaly.
Given a high quality measurement of $\Gamma (x)$, we can use the asymmetry
$$
A^{\pi^+ + \pi^-} =
{g_1^N(x) \over F_1(x)} - \Biggl( 1 - {D_4(z) \over (D_1 + D_2)(z)} \Biggr)
{1 \over 3} {\Gamma (x) \over F_1(x)}
\eqno(13)
$$
to extract the relative fragmentation of the anomaly
viz. $(1 - {D_4(z) \over (D_1+D_2)(z)})$.
Since the final state hadrons are dominated by light mass pions it is
reasonable
to take
$$
A^{\pi^+ + \pi^-} \simeq
A_1 \simeq {g_1^N(x) \over F_1^N (x)}
\eqno(14)
$$
whence
$$
D_4(z) = D_1(z) + D_2 (z)
\eqno(15)
$$
describes the fragmentation of the background field contribution.

As a guide to the size of the effect that one is looking for we evaluate
the
asymmetry in equ.(12)
using the phenomenological parametrisation of the anomaly in [27]
(which was deduced
via the
MIT bag model).
This is shown in Fig.1.
Here the dashed curve is the semi-inclusive asymmetry assuming zero large $x$
anomaly;
the bold curve is the model calculation with a large $x$ anomaly included.
(Given the semi-classical nature of the model and the non-classical nature of
the anomaly,
it is important to regard Fig.1 more as a guide to the experimental
accuracy
that is required
than a rigorous prediction of the asymmetry.)

In summary, semi-inclusive measurements of fast pions in polarised deep
inelastic
scattering
offer a window where we may hope to make an explicit measurement of the
background
field in QCD.
Given the programme of present [20,21]
and proposed [34]
spin experiments in this field
together with data from future high-energy, polarised, hadron hadron collisions
[4],
we may soon begin to learn
about this important non-perturbative physics.

\vspace{1.0cm}

{\bf Acknowledgements \\}

I thank A.W. Thomas and W. Melnitchouk for helpful discussions.
This work has been supported in part by the EC Programme ``Human Capital and
Mobility", Network ``Physics at High Energy Colliders", contract
CHRX-CT93-0357 (DG 12 COMA).

\pagebreak

\begin{center}{\bf References}\end{center}
\vspace{1.0cm}
\begin{enumerate}
\item
The EMC Collaboration,
J. Ashman et al., Phys. Lett. B206 (1988) 364, Nucl. Phys. B328 (1990) 1
\item
R. Windmolders, Int. J. Mod. Phys. A7 (1992) 639
\item\label{jpg}
S.D. Bass and A.W. Thomas, Prog. Part. Nucl. Phys. 33 (1994) 449 \\
G. Altarelli and G. Ridolfi, Nucl. Phys. B (Proc. Suppl.) 39B (1995) 106
\item
O. Nachtmann and A. Reiter, Z Physik C24 (1984) 283 \\
A. Brandenburg, O. Nachtmann and E. Mirkes, Z Physik C60 (1993) 697 \\
O. Nachtmann, to appear in Proc. ELFE Summer School, Cambridge 1995 (Editions
Frontieres
1995)
\item
R. L. Jaffe, MIT preprint MIT-CTP-2466 (1995), submitted to Phys. Lett. B
\item
H. Fritzsch, Phys. Lett. B256 (1991) 75
\item
S.D. Bass, Z Physik C55 (1992) 653, ibid C60 (1993) 343
\item
S.D. Bass, Phys. Lett. B342 (1995) 233
\item
R.L. Jaffe, Phys. Lett. B193 (1987) 101 \\
R.L. Jaffe and A. Manohar, Nucl. Phys. B337 (1990) 509.
\item
The NA-10 Collaboration, S. Falciano et al., Z Physik C31 (1986) 513 \\
The NA-10 Collaboration, M. Guanziroli et al., Z Physik C37 (1988) 545
\item
A.V. Efremov and O.V. Teryaev, Dubna Preprint E2-88-287 (1988) \\
G. Altarelli and G.G. Ross, Phys. Lett. B212 (1988) 391
\item
J. Schwinger, Phys. Rev. 82 (1951) 664
\item
R. Jackiw, in ``Current algebra and anomalies", eds. S. B. Treiman,
R. Jackiw, B. Zumino and E. Witten (World-Scientific, 1985)
\item
C. H. Llewellyn Smith, Oxford preprint OX-89/88 (1988) \\
C. H. Llewellyn Smith, Nucl. Phys. A434 (1985) 35c
\item
R. L. Jaffe and G. G. Ross, Phys. Lett. B93 (1980) 313 \\
R. L. Jaffe, Nucl. Phys. B229 (1983) 205
\item
A. W. Schreiber, A. W. Thomas and J. T. Londergan, Phys. Rev. D42 (1990) 2226
\\
A. W. Schreiber, A. I. Signal and A. W. Thomas, Phys. Rev. D44 (1991) 2653
\item
J. Kogut and L. Susskind, Phys. Rev. D11 (1974) 3594
\item
J. D. Sullivan, Phys. Rev. D5 (1972) 1732
\item
G. M. Shore and G. Veneziano, Nucl. Phys. B381 (1992) 23
\item
The HERMES Proposal, K. Coulter et al., DESY/PRC 90-1 (1990) \\
M. D{\"u}ren, Proc. Zuoz Summer School, PSI-Proceedings 94-01 (1994) 273
\item
W. Wislicki, SMC report, hep-ex 9405012 (1994)
\item
J. Kuti and V. Weisskopf, Phys. Rev. D4 (1971) 3418
\item
F. E. Close, Nucl. Phys. B80 (1974) 269
\item
R. Carlitz and J. Kaur, Phys. Rev. Lett. 38 (1977) 673
\item
A. Sch\"afer, Phys. Lett. B208 (1988) 175
\item
S.J. Brodsky, M. Burkardt and I. Schmidt, Nucl. Phys. B441 (1995) 197
\item
S.D. Bass and A.W. Thomas, Phys. Letts. B312 (1993) 345
\item
R. P. Feynman, Phys. Rev. Lett. 23 (1969) 1415
\item
L.L. Frankfurt et al., Phys. Lett. B320 (1989) 141
\item
F.E. Close and R.G. Milner, Phys. Rev. D44 (1991) 3691
\item
St. Gullenstern et al., Phys. Lett. B312 (1993) 166
\item
W. Melnitchouk, G. Piller and A. W. Thomas, Phys. Lett. B346 (1995) 165 \\
S. A. Kulagin, W. Melnitchouk, G. Piller and W. Weise,
Phys. Rev. C52 (1995) 932
\item
T. Gerhmann and W. J. Stirling, Z Physik C65 (1995) 461
\item
The HMC Collaboration, E. Nappi et al., Letter of Intent, CERN/SPSLC 95-27
(1995)
\end{enumerate}

\vspace{2.0cm}
\begin{center} {\bf Figure caption} \end{center}

\begin{flushleft}
Fig.1. \\
The deuteron asymmetry equ.(12) evaluated in the model of ref.[27].
The bold curve is
the spin asymmetry given a large $x$ anomaly contribution to $g_1$.
The dashed curve is the asymmetry with no large $x$ anomaly contribution.

\end{flushleft}

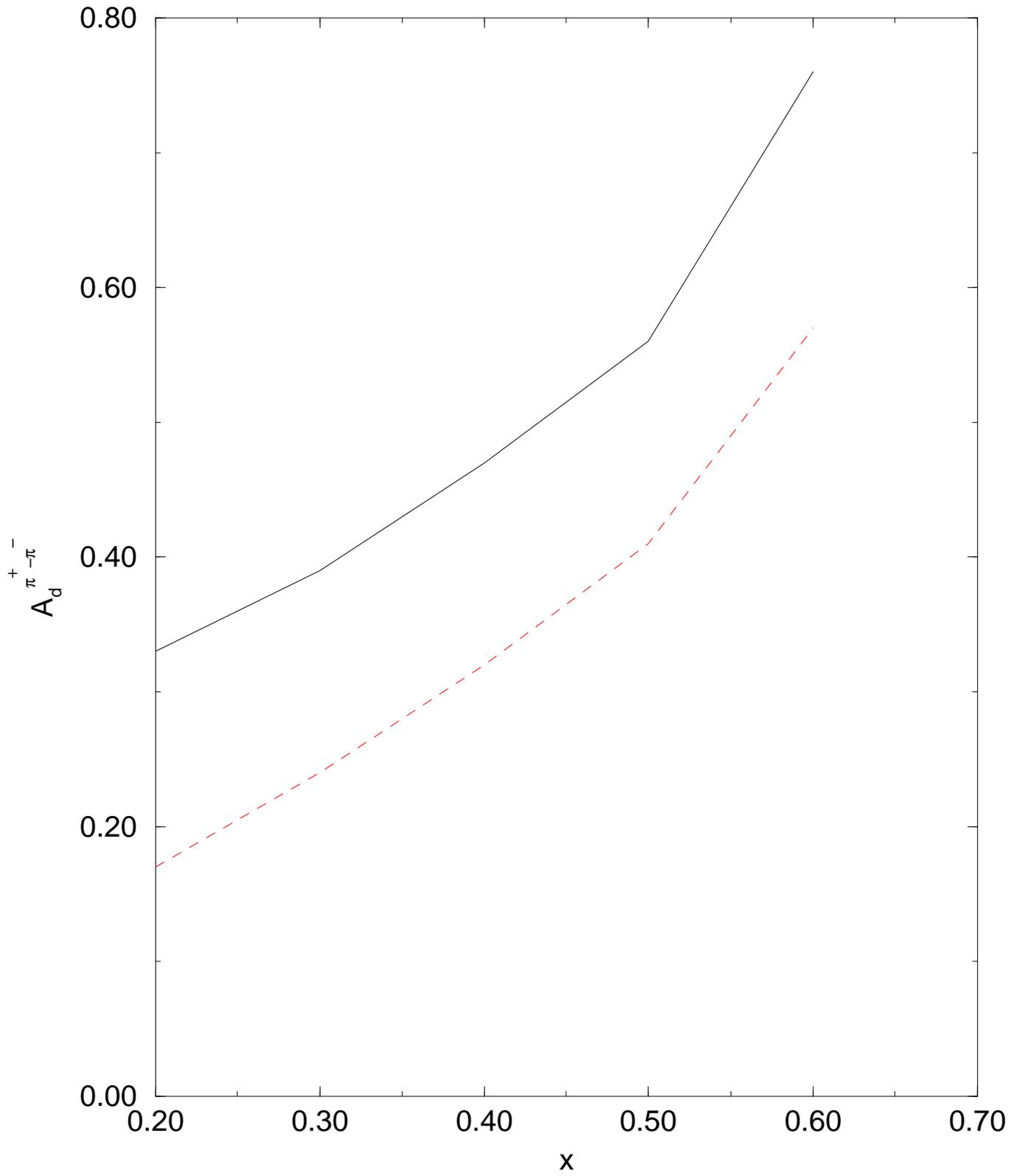
\end{document}